\documentclass[preprint2]{aastex}

\renewcommand\th{\thinspace}
\newcommand\kms{\ifmmode{\rm km\th s^{-1}}\else km\th s$^{-1}$\fi}
\newcommand\dg{$^{\circ}$}
\newcommand{\ha}{H$\alpha$}

\usepackage{graphicx}

\begin{document}

\title{A Search for Ionized Gas in the Draco and Ursa Minor Dwarf
  Spheroidal Galaxies}
\author{J.S. Gallagher, G. J. Madsen, R. J. 
Reynolds} \affil{Department of Astronomy, University of Wisconsin--Madison, 475 
  North Charter Street, Madison, WI 53706-1582; jsg@astro.wisc.edu, 
madsen@astro.wisc.edu, reynolds@astro.wisc.edu}
\author{E. K. Grebel}
\affil{Max-Planck Institute for Astronomy, K\"onigstuhl 17,                            D-69117 Heidelberg, Germany; grebel@mpia.de}
\author{T. A. Smecker-Hane} 
\affil{ Department of Physics \& Astronomy, 4129 Frederick Reines Hall, 
University of California, Irvine, CA 92697-4575; smecker@carina.ps.uci.edu}

\begin{abstract}
The Wisconsin \ha~Mapper has been used to set 
the first deep upper limits on the
intensity of diffuse \ha~emission from 
warm ionized gas in the Draco and Ursa Minor dwarf
spheroidal galaxies. Assuming a velocity dispersion of 15~km~s$^{-1}$
for the ionized gas,
we set limits of $I_{H\alpha} \leq 0.024$R and $I_{H\alpha} \leq
0.021$R for the Draco and Ursa Minor dSphs, respectively, averaged over our
1\dg~circular beam. Adopting a simple model for the ionized interstellar 
medium, these limits translate to upper bounds on the mass of ionized gas 
of $\lesssim$10\% of the stellar mass, or $\sim$10 times the 
upper limits for 
the mass of neutral hydrogen. Note that the Draco and Ursa Minor dSphs 
could contain 
substantial amounts of interstellar gas, equivalent to all of the gas 
injected by dying stars since the end of their main star forming 
episodes $\gtrsim$8~Gyr in the past, without violating these limits 
on the mass of ionized gas. 
\bigskip
\bigskip

\end{abstract}

\section{Introduction}

Among the unusual characteristics of dwarf spheroidal (dSph) galaxies
is the apparent lack of any interstellar gas, even in systems such as
Fornax or Carina, where star formation occurred in the last 1 to 3~Gyr
(Skillman \& Bender 1995; also see reviews by 
Gallagher \& Wyse 1994; Mateo 1998; Grebel 1999).
Stetson, Hesser, \& Smecker-Hane (1998) 
find a sequence of stars in color-magnitude 
diagram of the Fornax dSph that appears to be only a few times $10^8$ yrs 
old.  The
major exception to this trend may be the Sculptor dSph, where an
H\,{\sc i} cloud could be associated with the galaxy (Carignan
et al. 1998; Carignan 1999).  
Recent sensitive 21~cm H\,{\sc i} observations by Young
(1999, 2000) yielded upper limits on any diffuse neutral gas at column
densities of $< 10^{18}$~cm$^{-2}$, implying that any distributed gas
would have low mean volume density and therefore could be
ionized. Bowen et al. (1997) searched for a hot ionized ISM in the
outskirts of the Leo I dSph by seeking interstellar ultraviolet 
lines with no detection, thereby setting 
ionized gas column
density limits along two sight lines to background QSOs 
similar to those of the H\,{\sc i} surveys. Gizis, Mould and Djorgovsky (1993)
studied the Fornax dSph with {\it ROSAT} and found no extended component 
that could be associated with very hot gas and set a limit of 
$\leq 10^5$~M$_{\odot}$. 

The limits on interstellar gas column densities from 
the ultraviolet and H\,{\sc i} observations correspond to gas 
masses of $< 10^4$~M$_{\odot}$ or $\lesssim$1\% of the estimated stellar 
mass (see Grebel, Gallagher, \& Harbeck 2003). As noted by Young (2000), 
even if gas were initially removed from a dSph at the time that its star 
formation ceased, mass loss from stars should have replenished the 
interstellar medium to detectable levels. Evidently, dSphs either 
lose essentially all their interstellar gas  
or their interstellar gas lies in a form that has yet to be 
detected. For example, 
their gas could reside in dense, 
cold clumps or be nearly fully ionized but at a moderate 
temperature $\sim 10^4$~K that can be sustained by photoionization. 
In either case, 
substantial amounts of gas could 
escape detection by previous  H\,{\sc i}, x-ray,  
and ultraviolet absorption line studies.
In this paper we report results from observations with the highly-sensitive  
Wisconsin \ha~Mapper (WHAM) with the goal of detecting \ha~emission 
from ionized interstellar matter in the Draco and Ursa Minor dSphs.
No emission is found to sensitive limits, and we discuss the implications 
of these results. 

\section{Observations}

We have used WHAM to search for the
\ha~emission from the Draco and Ursa Minor dSphs. The WHAM instrument is a 
spectrometer located at the Kitt Peak National Observatory and is
remotely operated from Madison, Wisconsin.
WHAM consists of a 0.6 m siderostat coupled to a
15 cm dual-etalon Fabry-Perot system (Tufte 1997; Reynolds et al. 1998).
It produces an optical spectrum integrated over its circular, 1\dg\ diameter
field of view within a 
200 \kms~wide spectral window, centered
on any wavelength between 4800 \AA~and 7400 \AA, with a resolution of 12 \kms.
WHAM was designed to detect very faint emission lines from the diffuse
interstellar medium, and is capable of detecting \ha~emission down to
an emission measure of $\approx 0.02~$cm$^{-6}$ pc. It's capabilities are 
well-matched to surveys for faint diffuse \ha\ emission from extended 
sources, such as the Galactic satellite dSph galaxies.

We chose to study the Draco and Ursa Minor dSphs since they are the closest 
northern examples of these systems. 
Our WHAM observations 
were all taken on the night of
2002 May 6, and employed an 'ON minus OFF' technique. In order to
accurately remove the contamination of the spectra by faint
atmospheric lines (Madsen et al. 2001; Hausen et al. 2002), 
the observations were alternated 
between the galaxies (ONs), and a couple of directions
(OFFs) a few
degrees away from the ONs.  The OFF directions were
checked to ensure that no potential systematic contaminants, i.e. bright
stars or high velocity H\,{\sc i} 
gas, were present within the 200 \kms~spectral window.
Each individual observation was 120 s, and the
total integration time toward each ON direction was 78 min.
The average of the OFF spectra was subtracted from the average of the
corresponding ON spectra to yield spectra of the galaxies which
minimized the contamination by atmospheric lines.

Figures 1 and 2 show the resultant  \ha~spectra toward the Draco and Ursa
Minor dSphs, respectively. The spectra were corrected for atmospheric and
internal transmission effects, and their intensities calibrated using
synoptic observations of a portion of the North America Nebula, which
has an intensity $I_{H\alpha} = 800
~R$\footnote{$1~ \rm{R}=\frac{10^6}{4\pi}~$photons~cm$^{-2}$~sec$^{-1}$~sr $^{-1}$} 
(Scherb 1981).
The velocity scales were calibrated using a bright atmospheric OH 
line near -420 \kms with respect to \ha. The x-axis of each plot is 
heliocentric
velocity, and the y-axis is milli-Rayleighs per \kms.

\section{Results}

\subsection{Upper Limits on H$\alpha$ Intensities}

The spectra in Figures 1 and 2 show no statistically significant 
emission lines,
providing only upper limits to the \ha~emission from these
galaxies. To estimate these upper limits, we assume an emission line 
exists at
the heliocentric velocities of each of the dSph systems, -293 \kms~for
Draco and -248 \kms~for Ursa Minor (Armandroff, Olszewski, \& 
Pryor 1995). In order to set a
limit on the expected width of an \ha~line, we assume that the 
gas would have a velocity dispersion close to that of the  stars, 
which have line-of-sight dispersions of about 10
\kms (e.g.,  Olszewski, Aaronson, \& Hill 1995; Armandroff, Olszewski, \& Pryor 1995; Hargreaves et al 1996; Kleyna et al. 2002).  
However, it is likely that the velocity  
dispersion of the ionized gas would be larger than that of 
the stars, as the gas receives
additional energy input from the host stars as they
evolve. Furthermore, if we impose the condition that 
a significant amount of ionized gas is
actually present in these galaxies, we cannot use a dispersion that is
larger than is allowed in order to keep the gas bound in the system
($V_{esc} \gtrsim$ 15~\kms).
We therefore adopt a dispersion of 15 \kms~for the 
\ha~emission line that we attribute to ionized gas 
bound to the Draco and Ursa Minor dSph galaxies. 

Adopting the above heliocentric velocities and 15 \kms~dispersion, 
we use the observed spectra to
constrain the maximum area (intensity) of an undetected \ha\ line 
with these
parameters. In 
particular, we increase the area of the imposed 
spectra line until the average value of
the observed spectra within the FWHM of the line is 
more than 3$\sigma$ below 
the peak of the imposed line (See Figures 1 and 2). In this sense, we
consider the limits to the strengths of the imposed emission
lines to be 3$\sigma$ upper limits. We find upper limits of
$I_{H\alpha} \leq 0.024$R and $I_{H\alpha} \leq 0.021$R for 
the Draco and Ursa Minor dSphs, 
respectively, averaged over the 1\dg~circular WHAM beam. 
These limits are factors of 20-50 below the \ha~intensity limits 
obtained for external galaxies from deep, conventional 
narrow band filter imaging or 
spectroscopy (e.g., Ferguson et al. 1996; Hoopes, Walterbos, \& Rand 
1999), albeit at the price of the low angular resolution required 
to increase our sensitivity, 
but with the advantage of precision radial velocity information.

We note that these results were derived using
uncertainties arising only from the random errors (i.e. photon
statistics), whereas the systematic uncertainties from the incomplete
subtraction of atmospheric lines may also contribute to the overall
error measurements, particularly with the location of the
continuum. However, we have little understanding of the origin and
variability of these lines. Given that these random errors account for
point-to-point variation in the data, and 
that a straight line fit to the spectra, i.e. a spectrum with no 
features, yields a $\chi^2 = 1$, we
conclude that the contribution of systematic errors to our upper
limits is minimal. However, it is not clear if significantly 
deeper observations can be obtained with WHAM, as systematic 
errors are likely to become important for longer observations.

An additional consideration regarding these upper limits is
extinction of the
\ha~emission by interstellar dust in the Galaxy. However, 
according to the NASA Extragalactic Database, 
the predicted Galactic extinction at \ha~toward the Draco ($b^{II}=86 \degr$) 
and Ursa Minor ($b^{II}=45 \degr$) dSphs is less than 10\%. Furthermore, 
the variation in Galactic extinction across the projected angular 
extent of the Draco dSph is negligible (Odenkirchen et al. 2001) We do
not make any correction for Galactic extinction.

Another possible systematic 
error could be photospheric \ha~absorption by the stars in the host
galaxies themselves. This dip in the background stellar continuum would 
tend to reduce the measured intensity 
of any \ha\ emission feature from the ionized gas.
To estimate the magnitude of this effect, we consider a completely
saturated \ha~absorption line at the velocity of the host galaxy.
The strength of such a saturated line, as observed by WHAM, is at most
equal to the surface brightness, or continuum level, of the
galaxy. The apparent total $V$ magnitudes 
of the Draco and Ursa Minor dSphs, $m_V$, are 10.1 and 10.3, respectively
(Grebel et al 2003). In the limit where this light uniformly
fills the WHAM 1\dg~beam, we derive surface brightnesses of 0.7
$S_{10}$\footnote{$1~S_{10} = \rm{One~10th~mag~A0~star~deg}^{-2} =
  0.0044 \rm{R}\,/\,$\AA~at H$_{\alpha}$} and 0.5 $S_{10}$. Using a line
width of 50 \kms~(1 \AA), we find conservative upper limits of 
0.003 and 0.002 R for the strength of an \ha~absorption line. Given
that such a line is not likely to be completely saturated in these
systems, and that the line width at the core may be overestimated, it
is likely that any absorption line would be 
even weaker than these limits. We therefore have not taken into
account any correction for an internal stellar \ha~absorption
feature.

\subsection{Limits on the Mass of Ionized Gas}

With these upper limits of $I_{H\alpha}$ in hand, we estimate the
total mass of ionized gas in these galaxies. At a temperature of
$T=8000$ K, 1~R of \ha~emission corresponds to an emission measure of
$2.5~\rm{cm}^{-6}~\rm{pc}$. If we assume that the gas uniformly
fills a fraction $f$ of a cylinder of length $L$ along the line of
sight, we use the definition of emission measure $EM$: 
\begin{equation}
  EM =  f\,n_e^2\,L \approx 2.5\,I_{H\alpha}\,\epsilon^{-1}_{beam},
\end{equation}
where the beam efficiency $\epsilon_{beam}$ is given by the ratio of 
the circular 1\dg\ diameter WHAM field to the solid angle subtended 
by the source. The electron density is 
\begin{equation}
  n_e = EM^{\frac{1}{2}}\,f^{-\frac{1}{2}}\,L^{-\frac{1}{2}}.
\end{equation}
If we further assume that the gas only fills a sphere of radius
$R=\frac{L}{2}$ within this cylinder, then the geometric filling factor
is $f_g = \frac{2}{3}$, so that with an internal gas filling factor 
on small spatial scales of $f_i$, then the total filling 
factor is $f = f_i\,f_g $.

The mass of ionized gas $M_{ion}$ within a volume $V$ is given by
\begin{equation}
  M_{ion} = n_e\,m_H\,f\,C_Y\,V,
\end{equation}
with the mass of the hydrogen atom $m_H$, and the correction factor to include
helium $C_Y = 1.33$. Placing the emission within a sphere with 
$f_g = \frac{2}{3}$ and combining the above relations then yields
\begin{equation}
  M_{ion}=2 \pi\,\sqrt{\frac{5}{6}}\,m_H\,(f_i\,EM)^{\frac{1}{2}}\,R^{\frac{5}{2}} \end{equation}
$$= 0.059 (\frac{f_i}{0.1}\,I_{H \alpha}\,\epsilon^{-1}_{beam})^{\frac{1}{2}}\,R^{\frac{5}{2}}\,M_{\odot} , $$
where the radius $R$ is measured in parsecs.

To compute conservative limits we assume that the ionized gas, if present,
has a spherical distribution with radius equal to the tidal radius and 
an internal volume filling factor of $f_i = 0.1$ by analogy to conditions 
in the Galactic diffuse ionized gas (Reynolds 1977). 
This model is based on the assumption that any gas beyond the tidal 
radius would be lost (see Blitz \& Robishaw 2000). 
We adopt a tidal radius of $R_T = 34$~arcmin 
and a distance $D = 72$~kpc for Ursa Minor (Kleyna et al. 
1998; see also Carrera et al. 2002), 
and $R_T = 40$~arcmin with $D = 80$~kpc for Draco (Odenkirchen et al. 
2001). We then have $\epsilon_{beam} =$ 1 for both galaxies and  
$R_T =710 $~pc and 930~pc, respectively for the Ursa Minor and Draco dSphs.
Since we do not detect any emission, we do not correct for the 
elliptical shapes of these two galaxies, which could 
reduce $\epsilon_{beam}$ and increase in the upper mass limits 
by 20-25\%, if the gas had the same projected distribution as the stars 
but with uniform intensity.    
Incorporating these data into eq.(4) result in
3-$\sigma$ upper
limits to the ionized gas mass in Ursa Minor of $\le 1 \times
10^5 M_{\odot}$ and in Draco of $\le 2 \times 10^5 M_{\odot}$. If we instead 
assumed that the ionized gas is confined to the galaxies' core regions, 
these values would be reduced by about a factor of 5.

\subsection{Sources of Ionizing Radiation}

If \ha\ emission is present at its upper limit, we can assume ionization 
equilibrium to estimate the flux of
ionizing photons required to produce this amount of \ha~emission. For a
given $I_{H\alpha}$ in Rayleighs, the corresponding flux of Lyman
continuum photons $\phi$ incident on the surface of the galaxies is given by
\begin{equation}
  \phi = 2.05 \times 10^6\,I_{H\alpha}~\rm{[photons\,cm^{-2}\,s^{-1}]}
\end{equation}
where the factor of 2.05 comes from the number of Lyman continuum
photons need to produce one \ha~photon via photoionization
(Osterbrock 1989). The upper limit on $I_{H\alpha}$ derived for these two 
dSph galaxies would thus correspond to $\phi \approx 10^4$~photons 
cm$^2$~s$^{-1}$.
The total rate of Lyman continuum photons Q incident upon a galaxy
having a projected area A is then
\begin{equation}
  Q = \phi \, A~\rm{[photons\,s^{-1}].} 
\end{equation} For the projected areas
($\pi\, R_T^2$) of these galaxies, we thus find Lyman continuum photon
rates of $1.3 \times 10^{48}$~photons~s$^{-1}$ 
and $0.6 \times 10^{48}$~photons~s$^{-1}$ for the
Draco and Ursa Minor dSphs, respectively.

The hot stellar populations in the Draco and Ursa Minor dSphs, 
unfortunately, are poorly known. Given their dominant old stellar 
populations, hot stars arise from evolved low mass stars. For 
example, Ursa Minor has a predominantly blue horizontal branch 
(e.g., Kleyna et al. 1998; 
Bellazzini et al. 2002) while that in Draco the red horizontal branch is 
much stronger (Odenkirchen et al. 2001; Aparicio, A., Carrera, \& 
Mar\'tinez-Delgado 2001), and both galaxies 
should be producing white dwarfs. It is also possible that 
the Milky Way could ionize gas in these nearby dSphs. 
However, at the 70-80~kpc distances of these galaxies, it is unclear 
what the Milky Way's ionizing flux is. Detections of \ha\ emission 
from H\,{\sc I} clouds at distances of 10-50~kpc within the Galactic halo 
suggest an escaping flux  $\phi \approx 2 \times 10^5$~photons cm$^2$~s$^{-1}$
(Tufte et al. 1998; Bland-Hawthorne and Maloney 1999).  
Even the cosmic background may be sufficient to maintain ionization near 
our observed limit.
From WHAM \ha\ observations of the outer H\,{\sc i} disk of M31 
Madsen et al. (2001) set an upper limit of 
$\phi \leq 6 \times 10^4$~photons cm$^2$~s$^{-1}$ for the 
cosmic background in the Local Group. 
Note that assuming the galaxies are spheres, then 
$\phi$ as defined in our paper is 
four times larger than the $\Phi_0$ definition adopted by Madsen et al. 
Therefore, the cosmic background radiation field is sufficient 
to maintain an ionized 
ISM of substantial mass in the Draco and Ursa Minor dSphs even 
in the absence of stellar photoionization from 
the galaxies' themselves or from the Milky Way.

\section{Discussion and Conclusions}

Table 1 lists the estimated stellar masses and limits on gas masses for
the Draco and Ursa Minor dSphs. These results show that the H\,{\sc i}
limits now 
are at the level of $<$1\% of the stellar mass, while our new H\,{\sc ii}
results place limits of  order $\lesssim$ 10\% of the stellar mass.
Furthermore, we
find that current estimates of the background ionization rate would allow
gas masses near our limits to be maintained in an ionized state even
without additional energy inputs from the dSph stellar populations,
which are unlikely to be completely negligible (e.g., Burkert \&
Ruiz-LaPuente 1997).

Therefore, the old and extremely dim Ursa Minor and
Draco dSphs could harbor considerable amounts of ionized interstellar
matter. It is interesting to compare our limits with theoretical
expectations. Unfortunately, it is unclear how much gas should be left
over after early epochs of star formation in a small dwarf
galaxy. Several possibilities exist for the evolution of
interstellar matter in dwarf galaxies (e.g., Skillman \& Bender 
1995).  Currently available
theoretical models suggest that populations of cool gas in clouds can
survive star formation and associated supernovae over cosmic time
spans in dwarf galaxies, where galaxies with more dark 
matter should be better able to retain their gas 
(e.g. Hirashita 1999; Andersen \& Burkert
2000; Ferrara \& Tolstoy 2000)).  These models suggest that $>$10\% of
the baryonic mass remaining in low mass dwarfs that evolve over
$\sim$10~Gyr in isolation could be in the form of interstellar gas
(Lia, Carraro, \& Salucci 2000; Carraro et al. 2001). In this regard it 
is interesting to note that Ursa Minor and Draco may have different 
dark matter properties. Odenkirchen et al. (2001) find the outer regions 
of the Draco dSph has a regular structure, suggesting that perturbations 
induced by the Milky Way's tidal force are small leading to the 
conclusion that Draco is dominated by its dark matter halo. The Ursa Minor 
dSph, on the other hand, shows evidence for tidal disturbances 
(Mart\'{\i}nez-Delgado et al. 2001) and may have less dark matter, but also 
appears to be free of interstellar matter. 

A second option is that gas is removed via external processes. This may 
happen early on if reionization of the universe cuts off gas supplies 
to small galaxies, as suggested by some (Barkana \& Loeb 1999; 
Bullock, Kravtsov, \& Weinberg 2000; see also review by Loeb \& Barkana 
2001) but not all models (e.g., Ricotti, Gnedin \& Shull 2002). 
Evolution near giant 
galaxies provides additional opportunities for gas to be removed from 
dwarfs through ram pressure stripping or tidal effects (e.g., 
Sofue 1994; Grebel, 
Gallagher, \& Harbeck 2003).  Since the Ursa Minor and Draco 
dSphs are near the 
Milky Way, they probably experienced gas stripping.

Our observations in combination with the H\,{\sc i} data appear to 
rule out the possibility of a massive residual interstellar medium 
in either the Ursa Minor or Draco dSphs. We therefore consider how 
much gas might collect in these systems under the less strict assumption 
that they were cleaned of gas when their star formation ceased. 
Evolving stars are the most likely 
sources of interstellar gas for dSphs orbiting near the Milky 
Way. In an old stellar population the mass loss rate from 
stars is $ \dot M \approx 10^{-2} (\frac{5~Gyr}{t})$~(e.g., Schulz, Fritze-v. 
Alvensleben, \&  2002). Over 
an $\gtrsim$8~Gyr span appropriate to the time since star formation 
stopped in these two dwarfs (Carrera et al. 2002; 
Ikuta \& Arimoto 2002), we expect about 5\% of the stellar mass 
to return to the interstellar medium.  
A significant fraction of the gas 
returned by stars therefore cannot be in the form of H\,{\sc i}, but 
even with our new limits, this material could remain hidden in a diffuse 
ionized interstellar medium.

Of course, not all of the gas ejected by stars into the very empty 
interstellar space of a dwarf spheroidal system necessarily will 
remain in the galaxy. Ram pressure stripping can be highly 
effective if the orbits of dSphs take them within $<$ 50~kpc of  
the Milky Way (Sofue 1994; Grebel, Gallagher, \& Harbeck 2003). Thus 
the possibility remains that dSphs close to giants, like 
Ursa Minor and Draco, are nearly 
gas-free systems; our new observations do not disprove this 
option.

In summary, we obtained 1\dg\ angular resolution 
WHAM observations of the integrated \ha\ 
line fluxes from the Draco and Ursa Minor dSph companions of the 
Milky Way at their optical radial velocities. 
Our limits of $I_{H\alpha} < 0.024$R for Draco and 
$< 0.021$R for Ursa Minor are the most sensitive 
\ha\ emission line measurements for dSph galaxies to date. 
However, they still allow for the presence of a diffuse ionized 
interstellar medium with mass approaching 10\% of that of the 
stars, or $\sim$10 times that set by the H\,{\sc i} 21-cm line limits.  It is 
possible that even dSphs with dominant old stellar 
populations, like Draco and Ursa Minor, could contain 
the bulk of material lost by stars after star-formation activity
halted, or they 
may be periodically cleaned of gas during close 
encounters with the Milky Way. 
Searches for diffuse ionized gas in 
denser dSphs and dwarf galaxies further from the Milky Way will help 
to illuminate solutions to this long-standing mystery. To this end 
we recently obtained WHAM observations of the Fornax dSph a system 
that supported supported star formation, and will report these results 
in a subsquent paper.

\acknowledgments
JSG received support from the NSF through grant
AST98-03018, and from the University of Wisconsin Graduate School.
He also thanks Hans-Walter Rix for hospitality at the Max-Planck
Institute for Astronomy during the writing of this paper.
GJM and RJR acknowledge support from the NSF through grant
AST96-19424, and TSH from NSF grant AST00-70985. GJM also acknowledges 
support from the Wisconsin Space grant Consortium.  This
research has made use of the NASA/IPAC Extragalactic Database (NED)
which is operated by the Jet Propulsion Laboratory, California
Institute of Technology, under contract with the National Aeronautics
and Space Administration.

\begin{deluxetable}{lccc}
\tablecaption{Baryonic Mass in the Draco and Ursa Minor dSphs 
in Solar Masses} 
\tablewidth{0pt}
\tablehead{
\colhead{Galaxy} & \colhead{M$_* = 3L_V$} & \colhead{H\,{\sc i} Mass} & 
\colhead{H\,{\sc ii} Mass}
} 
\startdata
Draco      &  $1.4 \times 10^6$ & $<$ 8000   & $\le 2 \times 10^5$    \\
Ursa Minor &  $9 \times 10^5$   & $<$ 7000   & $\le 1 \times 10^5$     \\
\enddata
\tablecomments{Stellar luminosities and H\,{\sc I} masses from literature, 
as given by Grebel, Gallagher, \& Harbeck (2003). Ionized gas mass, 
M(H\,{\sc ii}), from this paper.}
\end{deluxetable}

\onecolumn

\begin{figure}[t]
  \includegraphics[scale=0.75,angle=90]{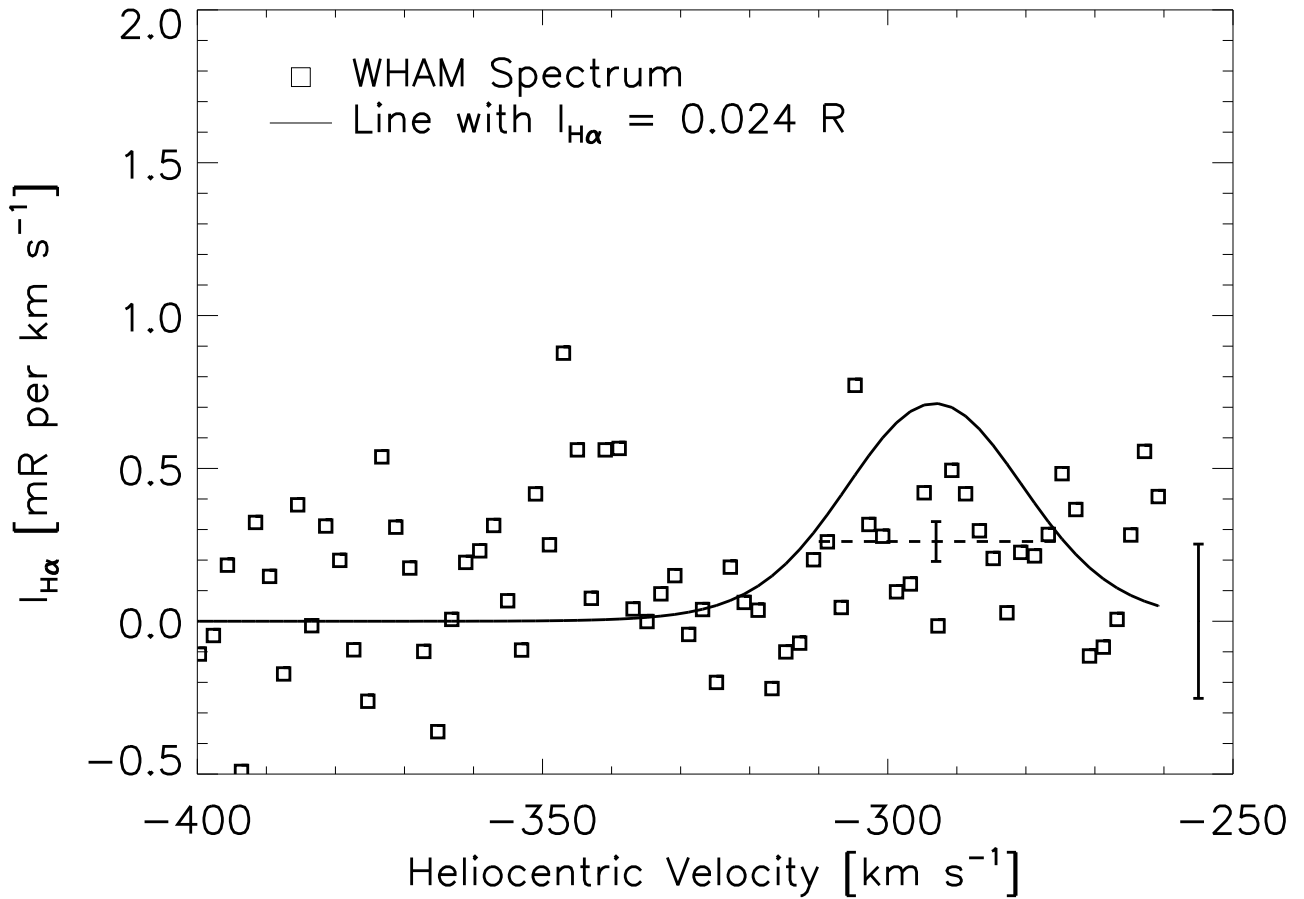}
  \caption{WHAM \ha~spectrum toward the Draco dwarf spheroidal galaxy. The 
    average of 34 OFF spectra were subtracted from the average of 34 ON
    spectra to produce this spectrum (see text). The x-axis is
    heliocentric velocity in \kms. The y-axis is
    \ha~intensity in milli-Rayleighs per \kms. A 
    characteristic 2-$\sigma_{data}$ error bar for each data point is shown at
    the far right near 
    -250 \kms. The solid line through the spectrum is a 3-$\sigma$ upper
    limit to an \ha~emission line, placed at the velocity of Draco with line
    width ($b$-value) of 15 \kms~and intensity $I_{H\alpha}=0.024
    \rm{R}$. The dashed line through middle of the 
    \ha~emission line is an average of all the data points that lie
    within the full width half maximum of the superimposed
    \ha~line. The error bar in the middle of this dashed line, 
    $\sigma_{ave}$, is the 
    average deviation of those data points from this line. The dashed
    line lies 3$\sigma_{ave}$ below the peak of the \ha~line. } 
  \label{fig:1}
\end{figure}

\begin{figure}[t]
  \includegraphics[scale=0.75,angle=90]{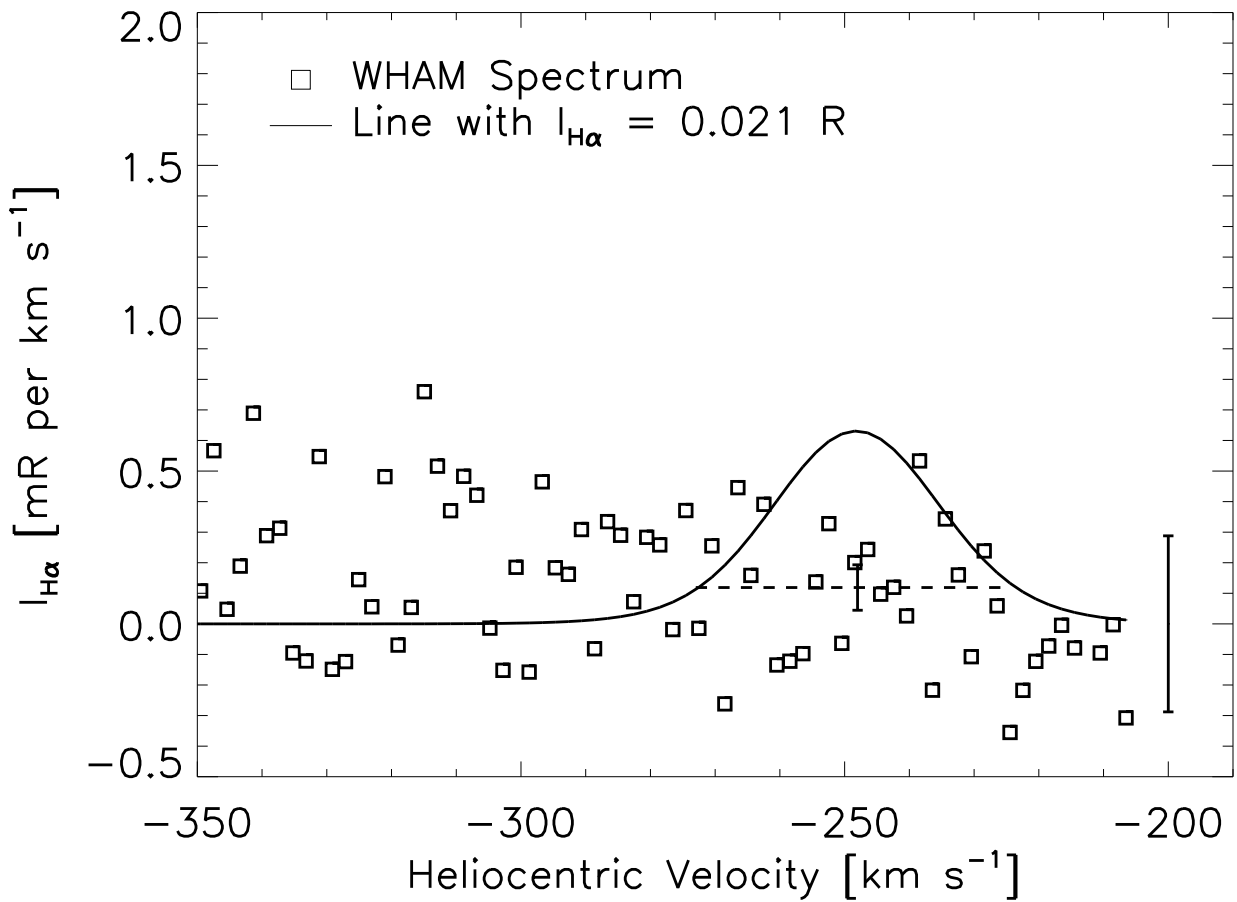}
  \caption{Same as Figure 1, but for the Ursa Minor dwarf spheroidal
    galaxy. The
    average of 16 OFF spectra were subtracted from the average of 32 ON
    spectra to produce this spectrum. A 
    characteristic 2-$\sigma_{data}$ error bar for each data point is shown at
    the far right near 
    -200 \kms. The solid line through the spectrum is a 3-$\sigma$ upper
    limit to an \ha~emission line, placed at the velocity of Ursa
    Minor with line 
    width ($b$-value) of 15 \kms~and intensity $I_{H\alpha}=0.021
    \rm{R}$.}
  \label{fig:2}
\end{figure}

\end{document}